\documentclass[12pt]{iopart}
\begin{document}

\title{New Nonrelativistic Quantum Theory of Cold Dark Matter}

\author{Z. E. Musielak}
\address{Department of Physics, The University of Texas at 
Arlington, Arlington, TX 76019, USA}
\ead{zmusielak@uta.edu}

\begin{abstract}
Cold dark matter (DM) is conceived as a gas of massive particles 
that undergo collisions, interact gravitationaly, and exchange quanta 
of energy.  A new nonrelativistic quantum theory is presented for 
this model of DM, based on a recently discovered equation for a 
spinless, no charge, and free particle.  This theory describes the 
quantum processes undergoing by the particles, specifies the 
required characteristic wavelength of the quanta of energy, gives 
constraints on the mass of DM particles, and predicts a detectable 
gravitational wave background associated with DM halos.
%
\end{abstract}


\section{Introduction}

The existence of dark matter (DM) is inferred from gravitational 
effects.  While the Universe contains 4.9\% of ordinary matter (OM),
data from the Planck 2018 mission [1] suggests that 26.8\% of DM 
constitutes the total mass-energy density.   

Although different theories of DM have been proposed, the physical 
nature of the particles that make up DM remains unknown [2-8]. 
The most commonly accepted theory proposes a weakly interacting 
massive particle (WIMP).  In the past several years, the search for 
WIMPs has intesified but thus far instruments have failed to detect 
them [8-10].  

In this paper, a new nonrelativistic quantum theory of DM is 
presented, based on an equation describing a spinless, without 
charge, and free particle [11] that may be considered a candidate 
for DM.  The existence of this equation is supported by the 
irreducible representations (irreps) of the extended Galilean group 
of the metric [12,13].  The irreps of the group also allow deriving 
the Schr\"odinger equation of quantum mechanics (QM) [14].  It 
was demonstrated that there is a special type of symmetry between 
these two equations and, since the Schr\"odinger equation describes 
the quantum structure of OM, the new equation may be used to 
represent DM [11].  

The preliminary quantum model of DM reported in [11] is significantly 
extended here by taking into account collisions between massive 
DM particles and their gravitational interaction, as well as exchange 
of quanta of energy.  The developed theory describes the quantum 
emission and absorption of the quanta of energy and the resulting 
equilibrium, and it also specifies the required characteristic wavelength 
of the quanta of energy, gives constraints on the mass of dark matter 
particles, and predicts a gravitational wave background for DM halos. 

This paper is organized as follows: the basic equations are given 
in Section 2; a model of dark matter is described in Section 3; a 
quantum theory of dark matter is presented in Section 4; physical
implications of the theory are discussed in Section 5; and Conclusions
are given in Section 6.   

\section{Basic equations of  nonrelativistic quantum physics}

In Galilean relativity, space and time are represented by different 
metrics that remain invariant with respect to all transformations 
that form the Galilean group of the metric $\mathcal {G} = [T(1) 
\otimes O(3)] \otimes_s [T(3) \otimes B(3)]$, where $T(1)$, $O(3)$, 
$T(3)$ and $B(3)$ are subgroups of translations in time, rotations, 
translations in space, and boosts, respectively [12].  However, the 
Schr\"odinger equation is invariant with respect to the extended 
Galilean group [15-17], whose structure is $\mathcal {G}_e = 
[O(3) \otimes_s B(3)] \otimes_s [T(3+1) \otimes U(1)]$, where 
$T(3+1)$ is an invariant Abelian subgroup of combined translations 
in space and time, and $U(1)$ is a one-parameter unitary subgroup 
[13].  

Classification of the irreps of $\mathcal {G}_e$ by Bargmann [18] 
demonstrated that only the scalar and spinor irreps are physical, 
but vectors and tensors are not because they do not allow for 
elementary particle localizations.  According to Wigner [19], a 
wave function must transform like one of the irreps of the group 
because, only in this case, all inertial observers identify the same 
physical object and agree on the description of its physical 
properties [20].  

It was shown [14] that the Wigner condition in the Galilean space 
and time can be expressed mathematically by two eigenvalue 
equations [14,17], which were used [11] to derive the following 
equations
\begin{equation}
\left [ i {{\partial} \over {\partial t}} + C_{s} \nabla^{2} \right ] 
\phi (t, \mathbf {x}) = 0\ ,
\label{eq1}
\end{equation}  
and 
\begin{equation}
\left [ {{\partial^2} \over {\partial t^2}} -  i  C_{w} {\mathbf k} \cdot 
\nabla \right ] \phi (t, \mathbf {x}) = 0\ .
\label{eq2}
\end{equation}  
where $\mathbf{x} = (x, y, z)$, with $x$, $y$ and $z$ being the Cartesian 
coordinates, $C_s = \omega / k^{2}$, and $C_w = \omega^2 / k^{2}$, 
with $k^{2n} = (\mathbf {k} \cdot \mathbf {k})^{n}$.  In addition, 
$\omega$ and ${\mathbf k}$ are labels of the irreps of $\mathcal {G}_e$, 
which means that they can be any real numbers.  As a result, there is an 
infinite number of equations given by Eqs (\ref{eq1}) and (\ref{eq2}). 

The origin of both equations is the same, namely, they are derived from
the eigenvalue equations that guarantee that the wavefunction $\phi (t, \mathbf {x})$ 
transforms as one of the irreps of $\mathcal {G}_e$.  The equations reflect 
properties of the Galilean spatial and temporal metrics, and they are the only 
second-order asymmetric differential equations allowed to be constructed in 
the Galilean space and time.  The equations complement each other and 
they form a set of twin-equations, whose physical applications are significantly 
different as demonstrated in this paper.

Since the metrics for space and time in Galilean relativity are separated, 
it is required that the derived equations are asymmetric with respect to 
the space and time derivatives, which is the necessary condition to make
the equations Galilean invariant.  Because of its form, Eq. (\ref{eq1}) is 
called a {\it Schr\"odinger-like equation}, while Eq. (\ref{eq2}) is referred 
to as a {\it new asymmetric equation} [11]. 

Among an infinite number of Schr\"odinger-like equations, the {\it Schr\"odinger 
equation} of QM can be obtained by specifying the constant $C_s$, which can 
be done when the de Broglie relationship is used [11].  The result is 
\begin{equation}
\left [ i {{\partial} \over {\partial t}} + \frac{\hbar}{2m} \nabla^{2} \right ] 
\phi (t, \mathbf {x}) = 0\ ,
\label{eq3}
\end{equation}  
which is the Schr\"odinger equation for a free particle.  By additing different 
potentials, the equation can be used to describe quantum states of OM in 
different physical settings [15].  

There are also infinitely many new asymmetric equations.  It is easy to verify
that the de Broglie relationship cannot be used to evaluate the constant $C_w$.
This means that $C_w$ does not depend on $\hbar$ but instead it requires a
new constant of Nature, denoted as $\varepsilon_o$, which represents a quanta 
of energy.  The resulting new asymmetric equation can be written as 
\begin{equation}
\left [ {{\partial^2} \over {\partial t^2}} -  i \frac{\varepsilon_o}{2m} 
{\mathbf k} \cdot \nabla \right ] \phi (t, \mathbf {x}) = 0\ ,
\label{eq4}
\end{equation}  
and it describes a free, spinless particle without charge.  There are 
differences between this equation and the Schr\"odinger equation as
Eq. (\ref{eq4}) has the second-order derivative in time instead of the 
first-order, and the first-order derivative in space instead of the second-order; 
thus, there is a special kind of space and time symmetry between these two 
equations.  In addition, the Schr\"odinger equation depends on the Planck 
constant that makes any quantized energy levels to depend on frequency, 
but the presence of quanta of energy $\varepsilon_o$ in the new asymmetric 
equation makes any quantized energy levels to be independent from frequency.    

Among, the main differences between the Schr\"odinger and the new asymmetric 
equations, the most important is that the latter cannot be made dependent on the 
Planck constant but instead it allows only for introducing the new constant 
$\varepsilon_o$.  This prevents the new asymmetric equation from being applicable
to any quantum description of OM.  Because of this limitation, it was already 
suggested that the new asymmetric equation may describe correctly the 
quantum structure of DM [11], with $\varepsilon_o$ being the quanta of energy 
of DM.  Since the new asymmetric equation does not depend on any potential
term, it may describe only a free particle of DM.  In the following, the equation 
is modified to account for the gravitational potential of DM, so the resulting new
asymmetric equation can describe gravitationally interacting DM.

\section{Dark matter model and governing equations}

Let $\mathcal{S}$ be a sphere of DM particles, or DM halo, and $m$ 
denotes the mass of each particle, which is charge and spin free.  For the 
particles uniformly distributed inside the halo, the gravitational potential 
$V_h (R)$ is given by 
\begin{equation}
V_h (R)  = \frac {G M_h}{2 R_h^3} \left ( R^2 - 3 R_h^2 \right )\ ,
\label{eq5}
\end{equation}  
where $M_h$ and $R_h$ are the mass and radius of the halo, and 
$R$ is the spherical coordinate [21].  The acceleration of DM particles
inside the halo resulting from this potential is
\begin{equation}
g_h (R) = \frac {dV_h (R)}{dR}  = \frac {G M_h}{R_h^3} R\ ,
\label{eq6}
\end{equation}  
which shows that $g_h (R)$ is a linear function of $R$ [21].  As 
a result, the derivative of $g_h (R)$ with respect to $R$ gives 
\begin{equation}
\Omega^2_h = \frac {dg_h (R)}{dR} = \frac {G M_h}{R_h^3} 
= {\rm const}\ ,
\label{eq7}
\end{equation}  
which shows that $\Omega^2_h$ remains the same at every point 
inside the halo.

A stable DM halo requires hydrostatic equailibrium, which means 
that 
\begin{equation}
\frac {dp_h (R)}{dR} = - \rho_h (R) g_h (R)\ ,
\label{eq8}
\end{equation}  
where $p_h (R)$ and $\rho_h (R)$ the pressure and density of DM 
inside the halo.  By writing the above equation in the form
\begin{equation}
g_h (R) = - \frac{1}{\rho_h} \frac {dp_h (R)}{dR}\ ,
\label{eq9}
\end{equation}  
the accelaration $g (R)$ becomes a pressure-gradient force per 
unit mass [22]. 

The particles of the spherical DM halo are allowed to collide, 
interact gravitationally, and also exchange energy by quantum 
processes.  To account for gravitational interaction between a 
pair of DM particles, a local spherical coordinate system $(r, 
\theta, \phi)$ centered at one of these particles is considered, 
with $r$ representing the distance between the particles.  In 
this model, the only spatial variable considered is $r$, which 
means that there are neither changes with respect to $\theta$
nor $\phi$. 

The gravitational potential $V (r)$ is the work that is needed 
to bring the DM particle from infinity to its location, and is 
given as $V (r) = - G m / r$.  By taking the negative gradient 
of $V (r)$, it yields the acceleration per unit mass of the particle, 
$a (r) = G m / r^2$ (see Eq. \ref{eq6}).   Taking the derivative 
with respect to $r$ one more time, the result is 
\begin{equation}
\Omega^2_{g} (r) = \frac{d a(r)}{dr} = - \frac {2 G m}{r^3}\ .
\label{eq10}
\end{equation}  
which can be included into Eq. (\ref{eq4}) to account for gravitational
interaction between a pair of DM particles.   Then, Eq. (\ref{eq4}) 
becomes 
\begin{equation}
\left [ {{\partial^2} \over {\partial t^2}} -  i \frac{\varepsilon_o}{2m} 
(\mathbf k \cdot \mathbf {\hat r}) \frac {\partial}{\partial r} - 
\Omega^2_g (r) \right ] \phi (t, r) = 0\ ,
\label{eq11}
\end{equation}  
with $\mathbf r = r \mathbf {\hat r}$.  Since $\Omega_h$ is constant 
anywhere inside the DM halo (see Eq. \ref{eq7}), the above new 
asymmetric equation with the $\Omega^2_g (r)$ term remains 
valid for any pair of particles located at any point inside the DM halo.  
This is an important result as it shows that the developed model of 
gravitationally interacting pair of DM particles is valid for the 
entire DM halo.  The model also allows for the particles of the 
pair to change their kinetic energy by colliding with other particles
in the DM halo.

According to [1-8,23-26], DM is not a source of any form of 
electromagnetic, weak or strong interactions, but it is known 
to interact gravitationally with OM.  However, in the model of 
DM considered in this paper, OM is not included, but only 
gravitationally interacting DM particles are taken into account.  

\section{Quantum theory of dark matter}

The quantum theory of DM is formulated by considering a pair of 
DM particles, whose evolution in time and space is described by 
Eq. (\ref{eq11}).  By separation of variables, $\phi (t, r) = \chi (t) 
\eta (r)$, the following equation is obtained
\begin{equation}
\frac {1}{\chi} \frac {d^2 \chi}{d t^2} = i \frac{\varepsilon_o}{2m} 
(\mathbf k \cdot \mathbf {\hat r}) \frac {1}{\eta} \frac {d \eta}{d r} + 
\frac {2 G m}{r^3} = - \mu\ ,
\label{eq12}
\end{equation}  
where $\mu$ is a separation constant to be determined.  For a 
time-independent model of DM, the equation to be solved is
\begin{equation}
\frac {d \eta}{\eta} = - 4 i \frac {G m^2}{\varepsilon_o}      
\frac {dr} {(\mathbf k \cdot \mathbf {\hat r}) r^3} - 2 i
\frac {m \mu}{\varepsilon_o} \frac {dr} {(\mathbf k \cdot 
\mathbf {\hat r})}\ ,
\label{eq13}
\end{equation}  
and the resulting solution is
\begin{equation}
\eta (r) = \eta_o \exp {\frac{2 i (\varepsilon_g - 2 \varepsilon_k)}
{\varepsilon_o (\mathbf k \cdot \mathbf r)}}\ ,
\label{eq14}
\end{equation}  
where $\eta_o$ is an integration constant, $\varepsilon_g = G m^2 / r$ 
is the gravitational potential energy of DM particles with mass $m$, and 
$\varepsilon_k = m v^2 / 2$ is the kinetic energy of the particles with 
their thermal velocity $v (r)$. The particle's thermal velocity is determined 
by the temperature of DM and rate of collisions between the particles 
in the DM halo. 

Let $\Delta \varepsilon = \varepsilon_g - 2 \varepsilon_k$, and $\eta_r (r)$
be the real part of the solution of $\eta (r)$ given by
\begin{equation}
\eta_r (r) = \eta_o \cos \left ( \frac{2 \Delta \varepsilon}{\varepsilon_o 
(\mathbf k \cdot \mathbf r)} \right )\ .
\label{eq15}
\end{equation}  
The following quantum processes are allowed:

(i) $\Delta \varepsilon = 0$ requires $\varepsilon_g = 2 \varepsilon_k$, 
which means that 
\begin{equation}
\cos \left ( \frac{2 \Delta \varepsilon}{\varepsilon_o (\mathbf k \cdot \mathbf r)} 
\right ) = 1\ ,
\label{eq16}
\end{equation}  
and that a pair of DM particles reaches its dynamical equilibrium, which is 
established between the gravitational potential and kinetic energies of the 
particles.  

In the equlibrium, the velocities of both particles of the pair are the same 
and given by $v_e = \sqrt {G m / r_e}$, where the subscript 'e' stands for 
equilibrium. Thus, the time $t_e$ required by a particle with its thermal 
velocity $v_e$ to travel the distance $r_e$ is $t_e = r_e / v_e$.  This allows 
finding the separation constant to be $\mu = 1 / t_e^2$ or $\mu = G m / 
r^3_e$, whose value is fixed.  Particles whose $v (r) \neq v_e$ travel in 
time $t_e$ the distance $r = v (r) t_e$. 

(ii) $\Delta \varepsilon > 0$ corresponds to emission of the quanta
$\varepsilon_o$ by the pair ($2 \Delta \varepsilon$).  However, the 
condition for the emission process for one DM particle of the pair is  
\begin{equation}
\frac{\Delta \varepsilon}{\varepsilon_o (\mathbf k \cdot \mathbf r)} 
= n \pi\ ,
\label{eq17}
\end{equation}  
with $n$ = 1, 2, 3, ..., as it guarantees that after this process takes 
place the pair reaches its dynamical equlilibrium.

(iii) $\Delta \varepsilon < 0$ represents absorption of the quanta 
$\varepsilon_o$ by the pair, and the condition for this process to 
occur for one DM particle of the pair is 
\begin{equation}
\frac{\Delta \varepsilon}{\varepsilon_o (\mathbf k \cdot \mathbf r)} 
= - n \pi\ ,
\label{eq18}
\end{equation}  
because after this process the pair returns to its dynamical equlilibrium.

In general, the quantization rules for one DM particle of the pair
can be written as 
\begin{equation}
\Delta \varepsilon = \pm n \pi \varepsilon_o (\mathbf k \cdot 
\mathbf r)\ ,
\label{eq19}
\end{equation}  
where $n$ = 0, 1, 2, 3, ..., with $n = 0$ representing the dynamical 
equilibrium of the pair.

A new result, when compared to QM, is the presence of the term 
$(\mathbf k \cdot \mathbf r)$, which shows that the emission and 
absorption processes depend on the direction between the vectors 
$\mathbf k$ and $\mathbf r$. The latter connects the particles of 
the pair but the former is the label of the irreps of $\mathcal {G}_e$ 
and it represents the inverse of the characteristic wavelength 
$\lambda_o$ associated with the quanta $\varepsilon_o$.  The 
wavelength is given by  
\begin{equation}
\lambda_o = \frac{G m^2}{\varepsilon_o}\ ,
\label{eq20}
\end{equation}  
with $\mathbf k = (1 / \lambda_o) \mathbf {\hat k}$, and plays the 
same role as the Compton wavelength $\lambda_c = h / m c$ in QM.  
However, while $\lambda_c$ depends on the Planck constant $h$ and 
the speed of light $c$, the characteristic wavelength for DM particles 
depends on the gravitational constant $G$ and on the new constant 
$\varepsilon_o$, implying that $\varepsilon_o$ may become
important in quantization of gravity, instead of the Planck constant.

The presented theory of quantum DM is nonrelativistic and based 
on a scalar wavefunction, as required by the irreps of $\mathcal {G}_e$; 
neither vector nor tensor wavefunctions can be used because they 
do not allow for localization of elementary particles [18].  Another 
nonrelativistic quantum theory based on the Schr\"odinger equation 
was proposed [27], and the theory postulated the existence of extremely
light bosonic particles; however, more detailed studies [28,29] showed 
that the theory failed as it required a DM particle of different mass for 
different types of galaxies.

Other theories of DM based on scalar [30,31] or other wavefunctions 
[32] are relativistic, developed within the framework of quantum field 
theory and typically they do not take gravitational interaction of DM into 
account.  Attempts to modify Einstein's General Relativty (GR) have been 
also made [33,34], but none of them is widely accepted because of the 
well-known predictive power of GR and its solid observational verification. 
Thus, the main difference between the previous theories of DM and the 
one developed in this paper is that the latter is nonrelativistic and based 
on the new asymmetric equation that accounts for gravitationally 
interacting DM.  

\section{Physical implications}

In the developed quantum theory of cold DM, massive particles 
are confined to a sphere (DM halo), and they form pairs bounded 
gravitationally.  The particles within the pairs may collide with 
other DM particles, changing the kinetic energy of the colliding 
particles.  A pair reaches its equlibrium when the gravitational 
potential energy equals the particle's kinetic energy.  If the 
gravitational energy exceeds the kinetic energy, a pair emits 
the quanta of energy $\varepsilon_o$; in the opposite case, 
a pair absorbs the same quanta.  This shows that $\varepsilon_o$
plays the same role for DM as the Planck constant does for OM.  

However, the main difference between $\hbar$ and $\varepsilon_o$
is that while the former is present in all quantum theories of OM 
that deal with electromagnetic, weak and strong interactions, 
the latter appears in the quantum theory of DM that deals with 
gravitational interaction, which is treated here classically.  
From a physical point of view, the situation resembles what is 
known in QM, where the electron described by the Schr\"odinger 
equation with the Coulomb potential requires photons to change 
orbits in the atom; however, nonrelativistic QM does not deal
with photons, which are introduced by relativistic QM or quantum
electrodynamics.  

Similarly, in the presented quantum theory of DM, the emission 
and absorption processes take place when the quanta $\varepsilon_o$ 
are exchanged.  Since the gravitational interaction is the only one 
considered in the theory, it is suggested that the quanta $\varepsilon_o$ 
are callled here {\it dark gravitons}, to emphasize that they are 
associated with DM.

Dark gravitons may be abundant in DM halos.  Such a large number 
of $\varepsilon_o$ in the sea of dark gravitos may be responsible for 
generating a gravitational wave background that would be specific for 
DM halos and, therefore, observable by the gravitational wave detectors 
as clearly distinct from the other proposed forms of stochastic gravitational 
wave background [26].   

The concept of massive gravitons has gained interest among 
researchers (e.g., [35-37]) because it can be used to account for DM, 
and depending on the graviton's mass, it may also be used to explain 
dark energy [36].  Having massive gravitons implies that gravitational 
waves do not propagate with the speed of light but their characteristic 
speed is lower.  The LIGO observations set up the limit on possible mass 
of gravitons as being smaller than $\approx 10^{-23}$ $eV / c^2$ [38].  
Dark gravitons introduced in this paper are distinst from either gravitons 
or masssive gravitons as the quantization procedure to obtain dark 
gravitons is significantly different than the quantum field procedure 
that gives gravitons or massive gravitons. 

Finally, it is interesting to consider the Compton wavelength 
$\lambda_c$ to be of the same order as the characteristic 
wavelength $\lambda_o$ of the quanta of $\varepsilon_o$ 
given by Eq. (\ref{eq20}).   More specifically, let $\lambda_c 
=\lambda_o$, which gives 
\begin{equation}
m = \sqrt[3] {\frac{h \varepsilon_o}{G c}}\ .
\label{eq21}
\end{equation}  
This shows that the mass of DM particles is fully determined by 
all four constants of Nature, with $\varepsilon_o$ being currently 
unknown.  However,  if Eq. (\ref{eq21}) is valid, then $\varepsilon_o$ 
would be known when the mass of DM particles could be experimentally 
established.  The four constants give the following characteristic 
length $l_c = \sqrt[3] {G h^2 / (\varepsilon_o c^2)}$.   

\section{Conclusions}

A spherical DM halo filled with massive DM particles is considered.
The particles are allowed to form gravitationally bounded pairs, 
whose dynamical stability can be impaired by particle collisions.
By using a recently discovered asymmetric equation for spinless,
no charge, and free particles [11], a quantum theory of cold DM 
is developed, showing that the dynamically unstable pairs can 
regain their stability by emitting or absorbing the quanta of energy 
$\varepsilon_o$. The quantum rules for these processes are 
presented and the characteristic wavelength corresponding to 
$\varepsilon_o$ is obtained.  Since the rules are independent from 
frequency, $\varepsilon_o$ is both a new constant of Nature as 
well as the quanta called dark gravitons associated exclusively 
with DM.  The abundance of $\varepsilon_o$ in DM halos forms 
the dark gravitons sea, which generate a gravitational wave 
background (GWB) that is specific for these halos and, therefore, 
potentially observable by the gravitaional wave detectors, as 
clearly distinct from other proposed forms of stochastic GWB [26].   

\bigskip\noindent
{\bf Acknowledgment:} The author is grateful to an anonymous 
referee for providing comments and suggestions that allowed 
significantly improve the revised version of this paper, and for 
bringing to my attention several papers on massive gravitons.
The author also thanks Dora Musielak for valuable comments 
on the earlier version of this manuscript.  This work was 
partially supported by Alexander von Humboldt Foundation.

\break

\end{document}